\documentclass[aps,showpacs,twocolumn,preprintnumbers,amsmath,amssymb]{revtex4}
\usepackage{graphicx}% Include figure files
\usepackage{bm}% bold math

\def\br{ \mathbf{r} }
\def\bR{ \mathbf{R} }
\def\bk{ \mathbf{k} }
\def\bq{ \mathbf{q} }
\def\bi{ \mathbf{i} }
\def\bj{ \mathbf{j} }
\def\bvs{ \mathbf{v}_s }

\def\Tr{ \,\mathrm{Tr}\, }
\def\tr{ \,\mathrm{tr}\, }

\begin{document}

\title{Nodal quasiparticles and classical phase fluctuations in $d$-wave superconductors}

\author{K.~V.~Samokhin and B.~Mitrovi\'c}

\address{Department of Physics, Brock University,
St.Catharines, Ontario, Canada L2S 3A1}
\date{\today}

\begin{abstract}
We show that the nodal quasiparticles have significant effect on
the classical phase fluctuations in a quasi-two-dimensional
$d$-wave superconductor. They give rise to singularities in the
temperature behavior of some of the coupling constants in the
phase-only effective action. One of the consequences is that the
classical $XY$-model is not adequate for the description of the
superconducting fluctuations in $d$-wave superconductors at low
temperatures.
\end{abstract}


\maketitle

In superconductors with low superfluid density $\rho_s$, the phase
fluctuations of the order parameter can play a significant role.
Emery and Kivelson \cite{EK95} suggested that it is the classical
phase fluctuations of the pre-formed Cooper pairs that are
responsible for the pseudogap phenomenon in the underdoped
high-$T_c$ materials (HTSC's) \cite{TS99}. This idea has been
further elaborated by many authors, see, e.g., the review article
\cite{LQS01}. On the other hand, it was argued in Refs.
\cite{RS95,CKEM99} that the phase fluctuations in HTSC's are
important down to $T\ll T_c$, which could explain a linear in $T$
behavior of the magnetic penetration depth $\lambda(T)$ at low
temperatures \cite{Hardy93}, even for a nodeless order parameter
(in a superconductor with lines of nodes, a linear $T$-dependence
of $\lambda$ occurs already at the mean-field level due to the
thermally excited nodal quasiparticles \cite{linearT}).

The effects of the phase fluctuations in HTSC's are usually
studied using either the classical or the quantum versions of the
$XY$-model, which is believed to provide a correct description of
any system with a one-component complex order parameter of fixed
magnitude. The energy of the classical two-dimensional (2D)
$XY$-model in the absence of external fields has the form
\begin{equation}
\label{XYdef}
    {\cal E}_{XY}=\sum\limits_{\br\br'}J_{\br\br'}
    [1-\cos(\theta_\br-\theta_{\br'})]
\end{equation}
Here $J_{\br\br'}$ are the phase stiffness coefficients, and
$\theta_\br$ is the phase of the order parameter at site $\br$ of
a coarse-grained square lattice. The lattice spacing $d$ is of the
order of the superconducting correlation length $\xi$, which is
almost constant at low temperatures. The $XY$-model has also been
applied to other systems, such as fabricated arrays of Josephson
junctions, or granular superconductors (see, e.g., Ref.
\cite{ES83} and the references therein), where $d$ is equal to the
distance between grains.

Although the simplicity of the model (\ref{XYdef}) is physically
appealing, it should be stressed that its rigorous microscopic
derivation for homogeneous high-$T_c$ superconductors does not
exist. The usual way of justification (see, e.g., Ref.
\cite{PRRM00}) involves expanding the cosine in the continuous
version of Eq. (\ref{XYdef}) and matching the coefficients with
those in the Gaussian phase-only action obtained by integrating
out the fermionic degrees of freedom. At low temperatures,
$\theta$ varies slowly at the length scale of $d$ and vortices can
be neglected. Then, assuming the nearest-neighbor interaction
$J_{\br\br'}=J$ in Eq. (\ref{XYdef}), introducing the superfluid
velocity $\bvs=\nabla\theta/2m$ ($m$ is the electron mass, and we
use such units that $\hbar=c=e=1$) and expanding the cosine in
powers of $\bvs$ and its gradients, we obtain
\begin{eqnarray}
\label{XYcont}
    {\cal E}_{XY}=\int d^2r\;\Bigl\{\frac{\rho_s}{2}\bvs^2+
    \frac{K}{2}\left[(\nabla_xv_{s,x})^2+(\nabla_yv_{s,y})^2\right]
    \nonumber\\
    +\frac{B}{4}\left(v_{s,x}^4+v_{s,y}^4\right)+...\Bigr\}.
\end{eqnarray}
Here $\rho_s=4m^2J$ is the mass density of the superfluid
electrons, $K=m^2d^2J$, and $B=-(8/3)m^4d^2J$. The ellipsis stands
for the terms of higher order in $\bvs$ or its gradients, such as
$(\nabla\bvs)^2\bvs^2$, {\em etc}. Given $J$ and $d$ in Eq.
(\ref{XYdef}), one can easily work out the coefficients in all
orders of the expansion. [If to take into account the
next-nearest-neighbor interactions $J_{\br\br'}$ \cite{KC02}, then
the tensor structure of the gradient and interaction terms in Eq.
(\ref{XYcont}) will change, for example the terms proportional to
$(\nabla_xv_{s,x})(\nabla_yv_{s,y})$, or $v_x^2v_y^2$ will
appear.] An important feature of the $XY$-model (\ref{XYdef}),
(\ref{XYcont}) is that $K$, $B$ and all the other expansion
coefficients are {\em temperature-independent} at low $T$.

That the consistent microscopic theory fails to reproduce the
quantum generalization of the $XY$-model has already been noticed
in Ref. \cite{BTC03}. In this Letter, we show that the
long-wavelength field theory even for the classical phase
fluctuations in a $d$-wave superconductor at low temperatures does
not have the form of Eq. (\ref{XYcont}). Briefly, the microscopic
theory shows that the coefficient in front of the gradient term in
the energy of fluctuations {\em diverges} at $T\to 0$, indicating
a strongly non-local interaction of the phase fluctuations. In
addition, the Taylor expansion of the energy of the phase
fluctuations in powers of the uniform superfluid velocity breaks
down at $T\to 0$, leading to a {\em non-analytical} dependence on
$\bvs$. To the best of our knowledge, the non-local and non-linear
effects on the phase fluctuations in high-$T_c$ superconductors
have not been emphasized before.

The starting point of our microscopic analysis is a tight-binding
Hamiltonian on a 2D square crystal lattice with the
``microscopic'' lattice constant $a$:
\begin{equation}
\label{Hamilt}
    H=\sum\limits_{\bi\bj}\xi_{\bi\bj}c^\dagger_{\bi\sigma}
    c_{\bj\sigma}-
    g\sum\limits_{\langle\bi\bj\rangle}B^\dagger_{\bi\bj}B_{\bi\bj}.
\end{equation}
Here $\xi_{\bi\bj}=-t_{\bi\bj}-\mu\delta_{\bi\bj}$, $t_{\bi\bj}$
is the hopping amplitude, $\mu$ is the chemical potential, $g>0$
is the superconducting coupling constant, and the operator
$B_{\bi\bj}=(c_{\bj\downarrow}
c_{\bi\uparrow}-c_{\bj\uparrow}c_{\bi\downarrow})/\sqrt{2}$
creates a singlet pair of electrons at the nearest-neighbor sites
$\langle\bi\bj\rangle$. The sites of the crystal lattice are
labelled by $\bi$, in contrast to the sites of the coarse-grained
lattice in Eq. (\ref{XYdef}). We assume zero external magnetic
field and neglect disorder and the electron-electron interactions
other than that responsible for the Cooper pairing (it can be
shown that the results below for the classical phase fluctuations
are not affected by the latter assumption).

Following the standard procedure, the partition function for the
model (\ref{Hamilt}) can be written as a Grassmann functional
integral. The interaction term are then decoupled using the
Hubbard-Stratonovich transformation, and the fermionic fields are
integrated out. After these manipulations \cite{details}, we end
up with a representation of the partition function as a functional
integral over a complex bosonic field $\Delta$ which is non-zero
only on the bonds between the nearest neighbors:
\begin{equation}
\label{FIDelta}
    Z=\int {\cal D}\Delta_{\bi\bj}(\tau){\cal
    D}\Delta^*_{\bi\bj}(\tau)\;e^{-S_{eff}[\Delta^*,\Delta]}.
\end{equation}
Here $0\leq\tau\leq\beta$ is the Matsubara time, $\beta=1/T$, and
the effective action is given by
\begin{equation}
\label{Seff}
    S_{eff}=-\Tr\ln {\cal G}^{-1}+\frac{2}{g}\int_0^\beta
    d\tau\;\sum\limits_{\langle\bi\bj\rangle}|\Delta_{\bi\bj}|^2,
\end{equation}
with
\begin{equation}
    {\cal G}_{\bi\bj}^{-1}=\left(\begin{array}{cc}
    -\delta_{\bi\bj}\partial_\tau-\xi_{\bi\bj} &
    \Delta_{\bi\bj}\\
    \Delta^*_{\bi\bj} & -\delta_{\bi\bj}\partial_\tau+\xi_{\bi\bj}
    \end{array}\right).
\end{equation}
We use the notation $\Tr$ for the full operator trace with respect
to the space-time coordinates and the Nambu indices, reserving the
notation $\tr$ for the $2\times 2$ matrix trace in the
electron-hole space.

The mean-field BCS theory corresponds to a stationary and uniform
saddle point of the functional integral (\ref{FIDelta}). The
$d$-wave saddle point in the momentum representation is given by
$\Delta_\bk=\Delta_0(T)\phi_\bk$, where $\phi_\bk=2(\cos k_xa-\cos
k_ya)$ is the symmetry factor. The temperature dependence of the
order parameter is determined by the standard BCS self-consistency
equation, generalized for the case of an anisotropic order
parameter (see, e.g., Ref. \cite{Book}). Deviations from the
mean-field solution can be represented in terms of the amplitude
and phase fluctuations:
$\Delta_{\bi\bj}(\tau)=[\Delta^{(0)}_{\bi\bj}+\delta\Delta_{\bi\bj}(\tau)]
e^{i\phi_{\bi\bj}(\tau)}$, where $\Delta^{(0)}_{\bi\bj}$ is the
Fourier transform of $\Delta_\bk$. We neglect the amplitude
fluctuations because they are gapped and therefore make a
negligible contribution at low temperatures. Since the number of
bonds in a square lattice is twice the number of sites, one should
introduce two phase fields on sites, $\theta_{\bi}(\tau)$ and
$\tilde\theta_{\bi}(\tau)$, to describe the phase degrees of
freedom. One possible parametrization is
$\phi_{\bi\bj}=\theta_\bi$ for $\br_\bj=\br_\bi+a\hat x$, and
$\phi_{\bi\bj}=\theta_\bi+\tilde\theta_\bi$ for
$\br_\bj=\br_\bi+a\hat y$. As shown in Ref. \cite{PRRM00}, the
fluctuations of $\tilde\theta$, which describe a change in the
symmetry of the order parameter from $d$-wave to $d+is$-wave, can
be neglected. Making a gauge transformation ${\cal
G}^{-1}_{\bi\bj}\to\tilde{\cal G}^{-1}_{\bi\bj}=U^\dagger_\bi{\cal
G}^{-1}_{\bi\bj}U_{\bj}$, where $U=\exp[-(i/2)\theta\tau_3]$
($\tau_3$ is the Pauli matrix in the electron-hole space)
\cite{about_U}, and expanding the $\Tr\ln {\cal G}^{-1}=\Tr\ln
\tilde {\cal G}^{-1}$ in powers of the small temporal and spatial
gradients of $\theta$, we obtain a phase-only effective action.

Motivated by the arguments of Emery and Kivelson that the phase
fluctuations in HTSC's are predominantly classical down to low
temperatures, we consider in this Letter only the classical case,
which corresponds to neglecting the $\tau$-dependence of $\theta$.
The effective action then becomes $S_{eff}=\beta({\cal E}_0+{\cal
E})$, where ${\cal E}_0$ is the mean-field energy, and ${\cal E}$
is the energy of phase fluctuations. In the Gaussian
approximation,
\begin{equation}
\label{E}
    {\cal E}=\frac{1}{2}\int d^2r_1d^2r_2\;
    {\cal K}_{ij}(\br_1-\br_2)
    v_{s,i}(\br_1)v_{s,j}(\br_2),
\end{equation}
where $i,j=x,y$, and the Fourier transform of the kernel ${\cal
K}(\bR)$ has the form
\begin{equation}
\label{Kij}
    {\cal K}_{ij}(\bq)=
    \rho_{s,0}\delta_{ij}-m^2\Pi_{ij}(\bq).
\end{equation}
To simplify the notations, from this point on we assume a
quadratic band dispersion $\xi_\bk=(k^2-k_F^2)/2m^*$ with the
effective mass $m^*$ equal to the electron mass $m$. In Eq.
(\ref{Kij}), $\rho_{s,0}=n_0m/a^2$ is the superfluid density at
zero temperature, $n_0$ is the average number of electrons per
site, and $\Pi_{ij}$ is the current-current correlator:
\begin{eqnarray}
\label{Pi}
    \Pi_{ij}(\bq)&=&-2T\sum\limits_n\int
    \frac{d^2k}{(2\pi)^2}\;
    v_i v_j\nonumber\\
    &&\times\tr[{\cal G}_0(\bk+\bq,\omega_n){\cal
    G}_0(\bk,\omega_n)].
\end{eqnarray}
Here ${\cal G}_0(\bk,\omega_n)=
(i\omega_n\tau_0-\xi_\bk\tau_3-\Delta_\bk\tau_1)^{-1}$ is the
mean-field matrix Green's function, $\omega_n=(2n+1)\pi T$ is the
fermionic Matsubara frequency, and $\mathbf{v}=\nabla_\bk\xi_\bk$
is the Fermi velocity. The energy of excitations,
$E_\bk=\sqrt{\xi_\bk^2+\Delta_\bk^2}$, vanishes at the four gap
nodes $\bk=\bk_n=k_F(\pm\hat x\pm\hat y)/\sqrt{2}$, $n=1\div 4$.

At low temperatures, only the nodal excitations with
$\bk=\bk_i+\delta\bk$ are important, which allows one to linearize
both the quasiparticle dispersion $\xi_\bk=v_F\delta k_\perp$ and
the order parameter $\Delta_\bk=v_\Delta\delta k_\parallel$, where
$\delta k_\perp$ and $\delta k_\parallel$ are the components of
$\delta\bk$ perpendicular and parallel to the Fermi surface. The
anisotropy ratio $v_F/v_\Delta$ depends on the material and the
doping level, e.g. $v_F/v_\Delta=14$ and $19$ in optimally doped
YBCO and Bi-2212, respectively \cite{ratio}. The fermionic
excitations have a conical spectrum: $E_\bk=\sqrt{v_F^2\delta
k_\perp^2+v_\Delta^2\delta k_\parallel^2}$ \cite{Lee93,LW97}.
Calculating the Matsubara sum in Eqs. (\ref{Kij}), we find
\begin{equation}
\label{Kgen}
    \begin{array}{l}
    {\cal K}_{xx}(\bq)={\cal K}_{yy}(\bq)=\rho_{s,0}-{\cal K}_+(\bq)\\
    \\
    {\cal K}_{xy}(\bq)={\cal K}_{yx}(\bq)=-{\cal K}_-(\bq),
    \end{array}
\end{equation}
where
\begin{equation}
\label{Kpm}
    {\cal K}_\pm(\bq)=\frac{m^2}{2\pi}\frac{v_F}{v_\Delta}T
    \left[s\left(\frac{\gamma_{1}}{T}\right)\pm
    s\left(\frac{\gamma_{2}}{T}\right)\right].
\end{equation}
Here
\begin{equation}
\label{Gamma12}
    \gamma_{1,2}(\bq)=\frac{1}{\sqrt{2}}\sqrt{v_F^2(q_x\pm q_y)^2+
    v_\Delta^2(q_x\mp q_y)^2}
\end{equation}
are the energies of the nodal quasiparticles with $\delta\bk=\bq$,
and $s(x)$ is a dimensionless scaling function, which is defined
by a rather cumbersome integral, whose asymptotic expansions at
large and small $x$ are
\begin{equation}
\label{asymp}
    s(x)=\left\{\begin{array}{cl}
     2\ln 2+x^2/24 &,\ \mbox{at }x\to 0\\
    \pi x/8 &,\ \mbox{at }x\to\infty.
    \end{array}\right.
\end{equation}
We would like to emphasize that the expressions (\ref{Kpm}) are
exact in the nodal approximation, i.e. for the conical
quasiparticle spectrum. In terms of $\bq$, the applicability
region of the nodal approximation is
$\gamma_{1,2}(\bq)\ll\Delta_0$.

At finite temperatures, the long-wavelength behavior of the kernel
${\cal K}(\bR)$ corresponds to the limit $\gamma_{1,2}(\bq)\ll T$.
In this case, according to Eqs. (\ref{Kgen}) and (\ref{asymp}),
the Fourier components ${\cal K}_{ij}(\bq)$ are analytical
functions of momentum, which can be written in the following form:
\begin{equation}
\label{KijT}
    \begin{array}{l}
    {\cal K}_{xx}(\bq)={\cal K}_{yy}(\bq)=\rho_{s}
    \left[1-\tilde\xi^2(q_x^2+q_y^2)\right]\\
    \\
    {\cal K}_{xy}(\bq)={\cal K}_{yx}(\bq)=-2\rho_{s}\tilde\xi^2q_xq_y,
    \end{array}
\end{equation}
where $\rho_s(T)=\rho_{s,0}-(2\ln 2/\pi)(m^2v_F/v_\Delta)T$ is the
mean-field superfluid mass density suppressed by the thermally
excited nodal quasiparticles \cite{LW97}. To leading order in
$(v_\Delta/v_F)^2\ll 1$, the characteristic length $\tilde\xi$ is
given by
\begin{equation}
\label{xiT}
    \tilde\xi(T)=\left(\frac{m^2v_F^3}{48\pi
    v_\Delta\rho_{s}T}\right)^{1/2}=
    C\left(\frac{T_c}{T}\right)^{1/2}\xi_0,
\end{equation}
where $T_c$ is the critical temperature, $\xi_0=v_F/2\pi T_c$ is
the BCS coherence length, and $C=(\pi
m^2T_cv_F/12\rho_{s}v_\Delta)^{1/2}$ is almost constant at low
$T$. For a circular Fermi surface and $T=0$,
$C=\sqrt{(\pi^2/12)(v_F/v_\Delta)(T_c/\omega_F)}\sim 1$.

The quadratic $\bq$-dependence (\ref{KijT}) implies that the
kernel ${\cal K}(\bR)$ in real space is proportional to
$\exp(-|\bR|/\tilde\xi)$. This behavior is similar to that of the
electromagnetic response function in conventional $s$-wave
superconductors, see e.g. Ref. \cite{Tink96}. An important
difference however is that the characteristic length of
fluctuations $\tilde\xi$ is now {\em temperature-dependent}:
$\tilde\xi(T)\sim T^{-1/2}$ at $T\to 0$. Qualitatively, this can
be attributed to the partial filling of the gap nodes by the
thermally-excited quasiparticles, which creates an effective
temperature-dependent $s$-wave gap. Note that the length
$\tilde\xi$, defined by the gradient expansion of the
current-current correlation function, Eq. (\ref{KijT}), is
different from other characteristic lengths discussed in the
literature: the coherence length $\xi_0$, which is the correlation
length of the amplitude fluctuations, and the size of a Cooper
pair $\xi_{pair}$, which is infinite in the $d$-wave case
\cite{BTCC02}. In a conventional $s$-wave superconductor, all
three lengths are of the same order.

If the typical length scale of the phase fluctuations exceeds
$\tilde\xi(T)$, then the energy of such fluctuations at finite
temperatures in the Gaussian approximation can be written as
${\cal E}=(1/2)(\int d^2R\,{\cal K}_{xx})(\int
d^2r\,\bvs^2)=(\rho_s/2)\int d^2r\,\bvs^2$, which indeed
reproduces the quadratic term in the continuum version of the
$XY$-model (\ref{XYcont}). However, comparing the gradient terms
in the $XY$-model (\ref{XYcont}) with the microscopic result
(\ref{KijT}), we see that they are clearly different. First, the
coefficient $K$ turns out to be {\em singular} at $T\to 0$.
Second, the sign and the tensor structure of the gradient terms
are different.

Because of the divergence of $\tilde\xi$, the gradient expansion
of ${\cal E}$ breaks down at $T\to 0$. As evident from Eqs.
(\ref{Kgen}) and (\ref{asymp}), at $T=0$ the matrix elements
${\cal K}_{ij}(\bq)$ are non-analytical functions of momentum. One
can therefore expect that the kernel ${\cal K}(\bR)$ decays
algebraically at large $\bR$. Indeed, calculating the Fourier
integrals, we obtain
\begin{equation}
\label{KzeroT}
    {\cal K}_\pm(\bR)\simeq\frac{m^2}{16\pi v_\Delta^2}
    [f_+(\bR)\pm f_-(\bR)],
\end{equation}
where
\begin{equation}
\label{fpmR}
    f_\pm(\bR)=\left[\left(\frac{R_x\pm R_y}{v_F}\right)^2+
    \left(\frac{R_x\mp R_y}{v_\Delta}\right)^2\right]^{-3/2}.
\end{equation}
Thus, the functions ${\cal K}_{ij}(\bR)$ are strongly anisotropic
and decay slowly, as $|\bR|^{-3}$, along a given direction. These
expressions also describe the behavior of ${\cal K}_{ij}(\bR)$ at
finite $T$ and at shorter distances, such that
$T\ll\gamma_{1,2}(\bq)\ll\Delta_0$. At larger distances there is a
crossover to the exponential decay obtained above.

In order to go beyond the Gaussian approximation, one has to take
into account the higher-order terms in the Taylor expansion of the
effective action (\ref{Seff}) in powers of $\nabla\theta$ or
$\bvs$. Assuming a uniform superfluid velocity, one can derive
from (\ref{Seff}) the following {\em exact} expression for the
energy density of fluctuations valid in all orders in $\bvs$
\cite{details}:
\begin{equation}
\label{lncosh}
    \frac{\cal E}{V}=-T\int\frac{d^2k}{(2\pi)^2}\;
    \ln\frac{\cosh\frac{\tilde E_{\bk,+}}{2T}
    \cosh\frac{\tilde E_{\bk,-}}{2T}}{\cosh^2\frac{\tilde
    E_{\bk}}{2T}},
\end{equation}
where $V$ is the system volume, $\tilde
E_{\bk,\pm}=(\xi_{\bk,+}-\xi_{\bk,-})/2\pm
\sqrt{(\xi_{\bk,+}+\xi_{\bk,-})^2/4+\Delta_\bk^2}$ are the
excitation energies affected by the superflow via the Doppler
shift in the energy spectrum, $\xi_{\bk,\pm}=\xi_{\bk\pm m\bvs}$.
Keeping the first terms in the Taylor expansion of ${\cal E}$, we
have
\begin{equation}
\label{Enl}
    \frac{\cal E}{V}=\frac{\rho_s}{2}\bvs^2+\frac{B}{4}
    \left(\bvs^4+4v_{s,x}^2v_{s,y}^2\right),
\end{equation}
where $\rho_s(T)$ is the mean-field superfluid density defined
above, and $B(T)={\rm const}-2m^2\tilde\xi^2(T)\rho_{s}\propto
T^{-1}$, to leading order in $T^{-1}$, with $\tilde\xi$ given by
Eq. (\ref{xiT}). We see that this expression does not have the
form of the continuum $XY$-model (\ref{XYcont}), since the
coupling constant $B$ is {\em divergent} at $T\to 0$. The
singularity of $B$ in Eq. (\ref{Enl}) indicates that the Taylor
expansion of the energy of fluctuations ${\cal E}$ in powers of
$\bvs$ breaks down at low temperatures. Indeed, at $T=0$ the
$\bk$-integral in Eq. (\ref{lncosh}) can be calculated, with the
result
\begin{eqnarray}
    \frac{\cal E}{V}=\frac{\rho_{s,0}}{2}\bvs^2-
    \frac{m^3v_F^2}{12\sqrt{2}\pi v_\Delta}\Bigl(
    |v_{s,x}+v_{s,y}|^3\nonumber\\
    +|v_{s,x}-v_{s,y}|^3\Bigr).
\end{eqnarray}
It is this non-analytical dependence that is responsible for the
non-linear Meissner effect in HTSC's \cite{non-lin} [to see this,
one should use the gauge-invariant superfluid velocity in the
presence of a vector potential $\mathbf{A}$,
$\bvs\to\bvs=(1/2m)\nabla\theta-\mathbf{A}$].

The non-linear terms in ${\cal E}$ are responsible for the
interaction of the thermally-excited classical phase fluctuations,
which in turn will lead to a renormalization of the bare
superfluid density $\rho_s(T)$. Despite the divergence of the
uniform coupling constant $B$ at $T\to 0$, it is possible to
calculate the lowest-order correction to $\rho_s$, which turns out
to be non-singular (and moreover it identically vanishes at $T=0$
if the quantum effects are taken into account \cite{BTC03}). The
formal reason is that the contribution of the uniform vertex
vanishes because of the momentum structure of the corresponding
diagrams.

In conclusion, we have shown that the effect of the $d$-wave nodal
quasiparticles on classical phase fluctuations cannot be ignored.
The microscopic derivation of a low-temperature effective energy
of the classical phase fluctuations gives a result that is very
different from the $XY$-model. The nodal quasiparticles with
conical dispersion lead to the appearance of a new,
temperature-dependent, length scale $\tilde\xi$, whose divergence
at $T\to 0$ causes both the gradient expansion and the Taylor
expansion in powers of $\bvs$ of the energy of fluctuations to
break down. The physical origin of the first effect is the same as
that of the non-local Meissner effect in $d$-wave superconductors
\cite{non-loc}: since the superconducting order parameter
$\Delta_\bk$ has nodes on the Fermi surface, then close to the
nodes anisotropic coherence length $v_F/|\Delta_\bk|$ exceeds the
London penetration depth $\lambda_0$, and the local
electrodynamics breaks down. The failure of the Taylor expansion
is closely related to the non-linear Meissner effect
\cite{non-lin}: the superfluid velocity acts as a pair-breaker in
$d$-wave superconductors, creating a finite density of normal
excitations even at $T=0$. The backflow of these excitations leads
to a non-analytical dependence of the electromagnetic response
functions on $\bvs$. The divergence of $\tilde\xi$ at $T\to 0$
might be cut off by the effects not included in our analysis, such
as quantum fluctuations, inter-layer tunnelling, or disorder.

The authors are pleased to thank J. Carbotte for pointing out Ref.
\cite{KC02} and E. Sternin for useful discussions. This work was
supported by the Natural Sciences and Engineering Research Council
of Canada.


\begin{thebibliography}{99}
\bibitem{details} The details of calculations will be published
separately.
\bibitem{about_U} Such choice of the transformation matrix $U$
results in the invariance under $\theta_\bi\to\theta_\bi+2\pi$
being lost. This does not lead to any consequences since we are
working in the low-temperature limit and the contribution of
vortices can be safely neglected.
\end{thebibliography}
\end{document}